\documentclass[12pt]{iopart}
\usepackage{iopams}
\expandafter\let\csname equation*\endcsname\relax
\expandafter\let\csname endequation*\endcsname\relax
\usepackage{amsmath}
\usepackage{amssymb}
\usepackage{mathrsfs}
\usepackage{type1cm}
\newcommand{\mathsfbi}[1]{\boldsymbol{\mathsf{#1}}}

\begin{document}

%\title[Author guidelines for IOP Publishing journals in  \LaTeXe]{How to prepare and submit an article for 
%publication in an IOP Publishing journal using \LaTeXe}
\title{Field Theory of Hyperfluid}

\author{Taketo Ariki${}^1$}

\address{${}^1$Institute of Industrial Science, The University of Tokyo, 4-6-1 Komaba, Meguro-ku, Tokyo 153-8505, Japan.}
%${}^2$Department of Physics, The University of Tokyo, 7-3-1 Hongo, Bunkyo-ku, Tokyo 113-0033, Japan}
\ead{${}^1$ariki@fluid.cse.nagoya-u.ac.jp}

\vspace{10pt}
\begin{indented}
\item[]May 2017
\end{indented}

\begin{abstract}
Hyperfluid model is constructed on the basis of its action entirely free from external constraints, regarding hyperfluid as a self-consistent classical field. Intrinsic hypermomentum is no longer a supplemental variable given by external constraints, but purely arises from the diffeomorphism covariance of dynamical field. Field-theoretic approach allows a natural classification of hyperfluid on the basis of its symmetry group and corresponding homogeneous space; scalar, spinor, vector, and tensor fluids are introduced as simple examples. Apart from phenomenological constraints, the theory predicts the hypermomentum exchange of fluid via field-theoretic interactions of various classes; fluid-fluid interactions, minimal and non-minimal $SU(n)$-gauge couplings, and coupling with the metric-affine gravity are all successfully formulated within the classical regime.
\end{abstract}

%\maketitle

\section{Introduction}
Fluid is typically employed as a macroscopic model of matters in the universe, providing one of fundamental pieces to spacetime structure via gravity-matter coupling. Increasing interests to go beyond the standard gravitational scenario motivate extensions of the known fluid concept. \emph{Spin fluid} in the torsion universe ($U_4$) of the Einstein-Cartan theory is one of such attempts, which had been studied on the line, at first, of the singularity avoidance or an alternative inflationary scenario of the early universe \cite{RS82,Gasp86,Obu87}. Gauge-theoretic reformulation of the spacetime geometry \cite{Hehl76,Hehl76-rev,Hehl95} encouraged scientists to go one-step further; the \emph{hyperfluid} \cite{Obu93}, fluid endowed with the hypermomentum, gives appropriate coupling charges to the metric-affine gravity. 

Although such a deep reconsideration from modern geometrical viewpoint appears strongly attractive to classical physics, the hyperfluid model still remains on the status of the phenomenology; fluid is no more than an approximation of the low-wavenumber mode of underlying manybody system, losing true benefit from the geometrical reconstructions of the spacetime. In conventional formalism \cite{Obu93,deRitis83,deRitis85,BF95,Obu96,BF98}, these phenomenological interpretations may be cast into constraints of two types. 
%In most of formalism \cite{Obu93,deRitis83,deRitis85,BF95,Obu96,BF98}, such a phenomenological interpretation is cast into constraints of two types. 
One is the Lin constraint \cite{Lin} which restricts the motion of coarse-grained properties such as mass, charge, or intrinsic hypermomentum onto the convection. The other is the Frenkel condition on the intrinsic hypermomentum, which further restricts the degrees of freedom originally allowed by the space-time symmetry, regarding the hyperfluid as assembly of microscopic objects. Whereas reflecting proper phenomenological interpretations, these conventional constraints strictly regulate the degrees of freedom, giving fluid's motion by external constraints rather than by internal structure of free Lagrangian. %In other words, effective-field interpretation is always lying behind the fluid concept, incorporating the intrinsic hypermomentum as a given variable whose motion is also given by several external constraints, which prevents hyperfluid from true reformulation based on the symmetry group. 
Indeed, there are careful observations by pioneers, where they recovered the lost symmetries by weakening the Frenkel condition \cite{BF95,Obu96,BF98}. Revisiting these facts, it may be naturally motivated, from the viewpoint of field theory, to explore the hyperfluid model totally based on symmetries, which would bring about the true extension of the known fluid concept at least in the classical regime, and the conventional hyperfluid models seem to be on a transient state. 
%Reminding the history of modern physics, symmetry-based extension often triggers the theoretical paradigm shift, and the conventional hyperfluid models seem to be on a transient state. 

The objective of this paper is to construct the theory of hyperfluid totally on the basis of symmetries inside its action. The resultant theory now defines the hyperfluid as a self-consistent classical field, where we do not rely on \emph{any} external constraints. Especially, the hypermomentum convection arises from the diffeomorphism covariance of action, regarding the diffeomorphism covariance of fluid as a gauge symmetry. Apart from phenomenological descriptions, the hyperfluid acquires the potential to have fundamental interactions; fluid-fluid coupling, $SU(n)$-gauge coupling with non-minimal interaction, and coupling with the metric-affine gravity are all successfully formulated.

\section{Action of fluid}
Our action principle has its roots in the formulation of Ref. \cite{AM16}. We begin with the flat spacetime ($R_4$) of the metric-signature $diag(+,-,-,-)$. Only a multiple-component field $\phi^I$ (index $I$ represents its components) and a vector $\mathbf{w}$ (Taub's current \cite{Taub59}) are sufficient field variables of our fluid. Here we define the Lagrangian density as a real function $P(\mathfrak{D}\phi^I,\mathcal{W})$ of $\mathfrak{D}\phi^I(\equiv w^\nu\partial_\nu\phi^I)$ and $\mathcal{W}(\equiv -w^\nu w_\nu/2)$. The Euler-Lagrange equations read
\begin{subequations}
\begin{align}
&\frac{\delta S}{\delta\phi^I}=-\left(\sigma_I w^{\mu}\right){}_{,\mu}=0,\label{phi eq.}\\
&\frac{\delta S}{\delta w^{\mu}}
=\sigma_I \phi^I_{,\mu}-\tau w_\mu=0,
\label{w eq.}
\end{align}
\end{subequations}
where $\sigma_I=\partial P/\partial(\mathfrak{D}\phi^I)$ and $\tau=\partial P/\partial \mathcal{W}$. Taking the gradient of $P$ and using Eqs. (\ref{phi eq.})-(\ref{w eq.}) yield
\begin{equation}
\left\{(\rho+P)u_{\mu}u^{\nu}\right\}_{,\nu}=P{}_{,\mu}\ ,
\label{Euler eq.}
\end{equation}
where the Lagrangian density $P$, a normalized vector $\mathbf{u}(\equiv \mathbf{w}/|\mathbf{w}|)$ ($u^{\alpha}u_{\alpha}=1$), and $\rho(\equiv -2\tau\mathcal{W}-P)$ can be identified with the pressure, velocity, and mass (Time-like condition $w^\nu w_\nu>0$ is fulfilled by defining the function $P(\mathfrak{D}\phi^I,\mathcal{W})$ in the domain $\mathcal{W}<0$). The action has the symmetry under the shift $\phi^I\mapsto\phi^I+const.$, which leads to the convection currents $|\mathbf{w}|\sigma_I u^\mu$ without imposing Lin constraints \cite{AM16}. The Euler equation (\ref{Euler eq.}) identically holds for arbitrary component number of $\phi^I$, which allows fluid to couple with various symmetry group of arbitrary dimensions. In this paper, we impose the symmetry under the diffeomorphism group, where the hypermomentum naturally appears as corresponding Noether charge.

\section{Global Symmetry}
Here we consider the symmetry of action under non-degenerating global transformation of the linear coordinate frame given by 
\begin{subequations}
\begin{align}
x^\mu&\mapsto x^{\tilde{\mu}}=x^\mu+\epsilon^\mu+\epsilon^\mu{}_\nu x^\nu,
\label{global trans. x}\\
\phi^I&\mapsto \phi^I+\epsilon^\mu{}_\nu(\Sigma_\mu{}^\nu){}^I{}_J\phi^J,
\label{global trans. phi}
\end{align}
\end{subequations}
where $\epsilon^\mu$ and $\epsilon^\mu{}_\nu$ are arbitrary infinitesimal parameters. Transformation (\ref{global trans. x}) forms the affine group in a global sense, where the Poincar$\acute{\textrm{e}}$ group is generated when $\epsilon^\mu{}_\nu$ is antisymmetric. $\mathsfbi{\Sigma}^{\mu\nu}$ is the transformation generator whose anti-symmetric part generates the Lorentz transformation. The remaining dilatation and shear transformations require the metric ($g_{\mu\nu}$) to \emph{globally} transform, so the action may be written as $S=\int P \sqrt{-g}d^4x$ with $g$$(\equiv\mathrm{det}[g_{\mu\nu}])$ being constant in spacetime. The symmetry under Eqs. (\ref{global trans. x})-(\ref{global trans. phi}) yields the corresponding two Noether currents; the canonical energy-momentum ($T^\mu{}_\nu$) and the \emph{total} hypermomentum which is split into \emph{orbital} ($T^{\rho\mu}x^\sigma$) and \emph{intrinsic} parts ($\Delta^{\rho\sigma\cdot\mu}$):
\begin{subequations}
\begin{align}
T^\mu{}_\nu&=\frac{\partial P}{\partial \phi^I_{,\mu}}\phi^I_{,\nu}-P\delta^\mu_\nu
=(\rho+P)u^\mu u_\nu-P\delta^\mu_\nu,\label{canonical energy-momentum}\\
\!\!\!\!\Delta^{\rho\sigma\cdot\mu}
&=-\frac{\partial P}{\partial \phi^I_{,\mu}}(\Sigma^{\sigma\rho}){}^I{}_J\phi^J
=-|\mathbf{w}|\sigma_I(\Sigma^{\sigma\rho}){}^I{}_J\phi^J u^\mu.
\label{hypermomentum current}
\end{align}
\end{subequations}
Eq. (\ref{hypermomentum current}) explicates that the fluid carries the hypermomentum density as a convective charge. Thus, the hyperfluid is obtained by imposing the global symmetry under Eq. (\ref{global trans. x}) which forms a subgroup of $Di\!f\!f(4,R)$. Also note that hypermomentum does not give any modification to $T^\mu{}_\nu$, yielding the Euler equation identical to Eq. (\ref{Euler eq.}). The Noether theorem yields the canonical energy-momentum conservation $T^\mu{}_{\nu,\mu}=0$ and the quasi-conservation of the hypermomentum $\Theta^{\rho\sigma}=T^{\rho\sigma}+\Delta^{\rho\sigma\cdot\mu}{}_{,\mu}$, where $\Theta^{\mu\nu}\equiv-2(\delta S/\delta g_{\mu\nu})/\sqrt{-g}$ (the \emph{metrical energy-momentum tensor}) generalizes the known Belinfante-Rosenfeld tensor \cite{Hehl76}. Following Ref. \cite{Hehl76} the hypermomentum current is split into the \emph{spin}, \emph{dilatation}, and \emph{shear currents} (traceless proper hypermomentum), which is given in this order by $S^{\rho\sigma\cdot\mu}\equiv\Delta^{[\sigma\rho]\cdot\mu}$, $\Delta^\mu=\Delta^\rho{}_\rho{}^\mu$, and $\bar{\Delta}^{\rho\sigma\cdot\mu}\equiv\Delta^{(\rho\sigma)\cdot\mu}-\frac{1}{4}\eta^{\rho\sigma}\Delta^{\mu}$.
\if0
\begin{subequations}
\begin{align}
S^{\rho\sigma\cdot\mu}&=-|\mathbf{w}|\sigma_I\left(\Sigma^{[\sigma\rho]}\right){}^I{}_J\phi^J u^\mu,\label{spin}\\
\Delta^\mu&=-|\mathbf{w}|\sigma_I(\Sigma^\rho{}_\rho){}^I{}_J\phi^J u^\mu,\label{dilatation}\\
\bar{\Delta}^{\rho\sigma\cdot\mu}&=-|\mathbf{w}|\sigma_I\left(\Sigma^{(\sigma\rho)}\right){}^I{}_J\phi^J u^\mu-\frac{1}{4} \eta^{\rho\sigma}\Delta^\mu.\label{shear}
\end{align}
\end{subequations}\fi
In case of the free hyperfluid, canonical energy-momentum tensor $T^{\mu\nu}$ is always symmetric as Eq. (\ref{canonical energy-momentum}), which yields, using the Belinfante-Rosenfeld relation, $S^{\rho\sigma\cdot\nu}{}_{,\nu}=0$, where the spin current independently conserves while the dilatation and shear currents may interact with the orbital part.

\section{Classifications}
Explicit form of Eq. (\ref{hypermomentum current}) varies with respect to each generator $\mathsfbi{\Sigma}^{\mu\nu}$, which allows us to classify the hyperfluid. If $\phi$ transforms as a scalar multiplet, the fluid has no intrinsic hypermomentum. We may call this as the \emph{scalar fluid}, which defines an ordinary fluid without hypermomentum. If $\phi$ transforms as a spinor (\emph{spinor fluid}), we write the Lagrangian density as $P(\Upsilon,\mathcal{W})$, where $\Upsilon\equiv (\mathfrak{D}\phi)^\dagger \gamma^0\mathfrak{D}\phi$ and its conjugate $\bar{\sigma}=\partial P/\partial(\mathfrak{D}\phi)=\frac{\partial P}{\partial\Upsilon} (\mathfrak{D}\phi)^\dagger \gamma^0$ (bar for the Pauli conjugate; $\bar{\sigma}\equiv \sigma^\dagger \gamma^0$). Substituting the generator $\Sigma^{\mu\nu}=\frac{1}{8}[\gamma^\mu,\gamma^\nu]$ into Eq. (\ref{hypermomentum current}) reads
\begin{equation}
\Delta^{\rho\sigma\cdot\mu}=\frac{1}{8}|\mathbf{w}|\left(\bar{\sigma}[\gamma^\rho,\gamma^\sigma]\phi-\bar{\phi}[\gamma^\rho,\gamma^\sigma]\sigma\right)u^\mu,
\label{spinor}
\end{equation}
which only contains the spin current \cite{manifield}; the spinor fluid defines a \emph{pure spin fluid} free from the dilatation and shear currents. Here $\gamma^\mu$ is given by $e^\mu_a\gamma^a$ using the Dirac matrices $\gamma^a$ in the orthonormal-frame representation $(g_{\mu\nu} e^\mu_ae^\nu_b=\eta^{ab})$, where the Greek indices stand for the coordinate (holonomic) components while the lowercase Latins for the orthonormal-frame (unholonomic) ones \cite{frame}. If $\phi^\mu$ transforms as a vector (\emph{vector fluid}), substituting the generator $(\Sigma^{\rho\sigma}){}^{\kappa}{}_{\lambda}=\eta^{\rho\kappa}\delta^\sigma_\lambda$  into Eq. (\ref{hypermomentum current}) yields %$\Upsilon=\mathfrak{D}\phi^\mu \mathfrak{D}\phi_\mu$
\begin{equation}
\Delta^{\rho\sigma\cdot\mu}=-|\mathbf{w}|\phi^\rho \sigma^\sigma u^\mu.
\label{vector}
\end{equation} 
which now contains the spin, dilatation, and shear currents. Further generalization is made by letting $\phi^{\mu\nu\cdots}$ to be a tensor (\emph{tensor fluid}), where Eq. (\ref{hypermomentum current}) reads %$(\Sigma^{\rho\sigma}){}^{\xi\zeta}{}_{\kappa\lambda}=\eta^{\xi\rho}\delta^\sigma_\kappa\delta^\zeta_\lambda+\eta^{\zeta\rho}\delta^\sigma_\lambda\delta^\xi_\kappa$ and $\Upsilon=\mathfrak{D}\phi^{\mu\nu} \mathfrak{D}\phi_{\mu\nu}$ yield
\begin{equation}
\begin{split}
\Delta^{\rho\sigma\cdot\mu}
=-|\mathbf{w}|(\phi^{\rho\alpha\beta\cdots}\sigma^\sigma{}_{\alpha\beta\cdots}
+\phi^{\alpha\rho\beta\cdots}\sigma_\alpha{}^\sigma{}_{\beta\cdots}
+\phi^{\alpha\beta\rho\cdots}\sigma_{\alpha\beta}{}^\sigma{}_{\cdots}+\cdots&) u^\mu.
\end{split}
\label{tensor}
\end{equation}
Unlike the conventional models \cite{RS82,Obu93,deRitis83,deRitis85,BF95,Obu96,BF98}, we do not need the dynamical equation for $\boldsymbol{\Delta}$, but $\boldsymbol{\Delta}$ is directly calculated from $\phi^I$ and $w^\mu$ obtained from their Euler-Lagrange equations (\ref{phi eq.})-(\ref{w eq.}).\\ %In any type of hyperfluid given above, $P(\Upsilon,\mathcal{W})$ has the discrete symmetry under transformation $\phi^I\mapsto -\phi^I$. In case of tensor fluids (including vector fluid), this allows $\boldsymbol{\phi}$ to be a pseudo tensor, which does not alter the energy-momontum tensor and the resultant Euler equation. \\
%In any type of hyperfluid given above, $P(\Upsilon,mathcal{W})$ has the discrete symmetry under transformations; $(\phi^I,\mathbf{w})\mapsto(\pm\phi^I,\pm\mathbf{w})$, allowing $(\phi^I,\mathbf{w})$ to be pseudo quantities. \\
%Further generalization may be possible by letting $\phi^I$ be arbitrary representation of $GL(4,R)$, so the \emph{manifield} of \cite{Hehl95,Neeman77} may cover all possible class of the hyperfluid.\\  

%Here we shall review the pioneering model of the spin fluid by Weyssenhoff, which has the spin current as an additional quantity. The Weyssenhoff fluid is an idealization of incoherent matter whose microscopic spin survives in macroscopic fluid description. In this model, the spin current is given as an assumption $S^{\rho\sigma\cdot\mu}=S^{\rho\sigma}u^\mu$ with the Frenkel condition $S^{\mu\nu}u_\nu=0$ imposed on the spin density $S^{\rho\sigma}$. Thus, it is emphasized here that we do not have to impose the hypermomentum as additional variable, but it is purely caused by the internal symmetry of the action (\ref{hyper fluid}) without any external constraints. Also note that the hypermomentum current is rather a secondary quantity derived by the dynamical variable $\phi^I$.\\

\section{Charged hyperfluid}\label{Charged hyperfluid}
Charged hyperfluid is defined by imposing $SU(n)$ symmetry on $\phi^I$, which is conducted by rewriting the Lagrangian density as $P(\Upsilon,\mathcal{W})$ with $\Upsilon\equiv(\mathfrak{D}\boldsymbol{\phi}_I)^\dagger\mathfrak{D}\boldsymbol{\phi}^I$. Local symmetry requires $\mathfrak{D}\boldsymbol{\phi}^I\equiv w^\nu\mathcal{D}_\nu \boldsymbol{\phi}^I$, where $\mathcal{D}_{\mu}\equiv\partial_{\mu}-ie\mathsfbi{t}_aA^a_{\mu}$ is the gauge-covariant derivative ($e$, $\mathsfbi{t}_a$, and $A^a_{\mu}$ are the coupling constant, generator, and gauge field). Now $\boldsymbol{\phi}^I$ is the $SU(n)$ multiplet, where the index $I$ is left to represent the components of either spinor, vector, or tensor. Symmetries under $SU(n)$ and Eqs. (\ref{global trans. x})-(\ref{global trans. phi}) yield the following two Noether currents:
\begin{subequations}
\begin{align}
J^\mu_a &= -ie|\mathbf{w}|\left(\boldsymbol{\sigma}^\dagger_J\mathsfbi{t}_a\boldsymbol{\phi}^J-\boldsymbol{\phi}^\dagger_J\mathsfbi{t}_a\boldsymbol{\sigma}^J\right)u^\mu,
\label{charge current}\\
\!\!\!\!\Delta^{\rho\sigma\cdot\mu}
=&-\!|\mathbf{w}|\!\left(\boldsymbol{\sigma}^\dagger_I(\Sigma^{\sigma\rho}){}^I{}_J\boldsymbol{\phi}^J+\boldsymbol{\phi}^{\dagger J}(\Sigma^{\sigma\rho}){}^I{}_J\boldsymbol{\sigma}_I\right)\!u^\mu,
\label{charged-hypermomentum current}
\end{align}
\end{subequations}
which show that the fluid carries both charge and hypermomentum. The Noether identities read $\mathcal{D}_\nu J^\nu_a=0$ and $\Delta^{\rho\sigma\cdot\nu}{}_{,\nu}=T^{\rho\sigma}_\mathcal{F}-\Theta_\mathcal{F}^{\rho\sigma}$, where $\boldsymbol{\Theta}_\mathcal{F}$ and $\mathsfbi{T}_\mathcal{F}$ are metrical and canonical energy-momentum tensors of fluid, so quasi-conservation of $\boldsymbol{\Delta}$ is closed in the fluid; \emph{charged hyperfluid of the minimal-coupling does not exchange its total hypermomentum with that of gauge field}. Especially, since $\boldsymbol{\Theta}_\mathcal{F}$ and $\mathsfbi{T}_\mathcal{F}$ are both symmetric tensors, fluid's spin conserves thereby $S^{\rho\sigma\cdot\nu}{}_{,\nu}=0$; \emph{the minimally-charged hyperfluid is free from spin-orbit interaction}. Thus, the spin-orbit interaction requires the non-minimal coupling; the prototypical sector $\propto F^{\mu\nu}S_{\mu\nu}$ of Refs. \cite{Obu87,deRitis85,KN14,Nair16} expresses the magnetic-moment proportional to the spin density, while such an interpretation is valid only when the hyperfluid is constituted of spinning dusts or Dirac Fermions. In the present model, we do not presuppose such underlying system, but the hypermomentum (\ref{hypermomentum current}) purely arises from the symmetry of field $\boldsymbol{\phi}^I$ itself, and thus the conventional spin-orbit interaction does not necessarily hold. Also note that the conventional $S_{\mu\nu}$ cannot couple with the non-Abelian field strength $F^{\mu\nu}_a$ because of its lack of the generator index $a$. 

Let us consider an alternative non-minimal coupling which does not contain the derivative term $\mathcal{D}_\nu\boldsymbol{\phi}^I$ of the minimal coupling. In order to make a proper coupling with the field strength $F^{\mu\nu}_a$, we need a real-valued tensor, say $\chi^a_{\mu\nu} (=-\chi^a_{\nu\mu})$, composed of our dynamical variables $\boldsymbol{\phi}^I$. The simplest composition may be
\begin{equation}
\chi^a_{\mu\nu}=
\begin{cases}
\bar{\boldsymbol{\phi}}[\gamma_\mu,\gamma_\nu]
\mathsfbi{t}^a \boldsymbol{\phi}\ \ &\textrm{spinor fluid},\\
\frac{i}{2}(\boldsymbol{\phi}^\dagger_\mu \mathsfbi{t}^a\boldsymbol{\phi}_\nu-\boldsymbol{\phi}^\dagger_\nu \mathsfbi{t}^a\boldsymbol{\phi}_\mu)\ \ &\textrm{vector fluid},\\
\frac{i}{2}(\boldsymbol{\phi}^\dagger_{\mu\alpha\beta\cdots} \mathsfbi{t}^a\boldsymbol{\phi}_\nu{}^{\alpha\beta\cdots}-\boldsymbol{\phi}^\dagger_{\nu\alpha\beta\cdots} \mathsfbi{t}^a\boldsymbol{\phi}_\mu{}^{\alpha\beta\cdots})\ \ &\textrm{tensor fluid}.
\end{cases}
\label{magnetic moment}
\end{equation}
%\begin{equation}
%\chi^a_{\mu\nu}=
%\begin{cases}
%\mathfrak{D}\bar{\boldsymbol{\phi}}[\gamma_\mu,\gamma_\nu]
%\mathsfbi{t}^a\, \mathfrak{D}\boldsymbol{\phi}\ \ &\textrm{spinor fluid}\\
%\frac{i}{2}(\mathfrak{D}\boldsymbol{\phi}^\dagger_\mu\, \mathsfbi{t}^a\,\mathfrak{D}\boldsymbol{\phi}_\nu-\mathfrak{D}\boldsymbol{\phi}^\dagger_\nu \,\mathsfbi{t}^a\,\mathfrak{D}\boldsymbol{\phi}_\mu)\ \ &\textrm{vector fluid}
%\end{cases}
%\end{equation}
Then we may have the following model:
\begin{equation}
\mathcal{L}=P(\Upsilon,\mathcal{W})
-\frac{1}{4}F^{\mu\nu}_aF_{\mu\nu}^a+\xi\chi_{\mu\nu}^aF^{\mu\nu}_a, 
\label{magnetized hyperfluid}
\end{equation}
where $\chi^a_{\mu\nu}$ is interpreted as the magnetic moment based on $SU(n)$ symmetry, $\xi$ is a coupling constant. In this model, fluid's intrinsic hypermomentum is still given by Eq. (\ref{charged-hypermomentum current}). Then $\Delta^{\rho\sigma\cdot\mu}$ satisfies
\begin{equation}
\Delta^{\rho\sigma\cdot\nu}{}_{,\nu}+T_\mathcal{F}^{\rho\sigma}-\Theta_\mathcal{F}^{\rho\sigma}
=2\xi\chi^\rho{}_{\lambda}{}^a F^{\lambda\sigma}_a\neq 0,
\label{hypermomentum non-conversion}
\end{equation}
where non-vanishing right-hand side acts as the source or sink of the hypermomentum, representing the exchange of the intrinsic hypermomentum with the Yang-Mills field; not only spin, but also the dilatation and shear currents now couples to Yang-Mills field via the magnetic moment $\chi_{\mu\nu}^a$. Taking the antisymmetric part yields
\begin{equation}
S^{\rho\sigma\cdot\mu}{}_{,\mu}=\xi(F^{\alpha\rho}_a\chi_\alpha{}^{\sigma a}
-F^{\alpha\sigma}_a\chi_\alpha{}^{\rho a}),
\end{equation}
where fluid's spin does not conserve because of the $SU(n)$-magnetic moment.

\section{Fluid-fluid interaction}
Regarding $\phi^I$ as the fundamental field variable of the theory, we introduce interactions between hyperfluids on the basis of it. Consider two spinor fluids $({}^{{}_1}\!\phi,{}^{{}_1}\!\mathbf{w})$ and $({}^{{}_2}\!\phi,{}^{{}_2}\!\mathbf{w})$ with a simple coupling:
\begin{equation}
\mathcal{L}={}^{{}_1}\!P+{}^{{}_2}\!P+\xi({}^{{}_1}\!\bar{\phi}\,{}^{{}_2}\!\phi+{}^{{}_2}\!\bar{\phi}\,{}^{{}_1}\!\phi),
\label{2s model}
\end{equation}
where ${}^{{}_1}P$ and ${}^{{}_2}\!P$ are the sectors of fluids 1 and 2, respectively. Using Eq. (\ref{spinor}) and field equation $\delta S/\delta\, {}^{{}_1}\!\phi=0$ yields
\begin{equation}
{}^{{}_1}\!S^{\rho\sigma\cdot\nu}{}_{,\nu}=\frac{1}{8}\xi\left({}^{{}_2}\!\bar{\phi}[\gamma^\rho,\gamma^\sigma]{}^{{}_1}\!\phi-{}^{{}_1}\!\bar{\phi}[\gamma^\rho,\gamma^\sigma]{}^{{}_2}\!\phi\right),
\label{spin non-conversion}
\end{equation}
where spin of fluid 1 does not conserve. Using anti-symmetry in the labels $1$ and $2$, we immediately reach the total-spin conservation: $\partial_\nu({}^{{}_1}\!S^{\rho\sigma\cdot\nu}+{}^{{}_2}\!S^{\rho\sigma\cdot\nu})=0$. For coupling with the other hyperfluid, one may introduce the vector $j^\mu\equiv\bar{\phi}\gamma^\mu \phi$ (or axial vector ${}^{{}_A}\!j^\mu\equiv\bar{\phi}\gamma^5\gamma^\mu \phi$) in a similar manner to that of the fermion current, which allows coupling to vector and tensor fluids via index contraction; for instance, one may construct a coupling model of the spinor fluid $({}^{{}_s}\!\phi,{}^{{}_s}\!\mathbf{w})$, vector fluid $({}^{{}_v}\!\boldsymbol{\phi},{}^{{}_v}\!\mathbf{w})$, and 2-rank tensor fluid $({}^{{}_t}\!\boldsymbol{\phi},{}^{{}_t}\!\mathbf{w})$ given by
\begin{equation}
\mathcal{L}={}^{{}_s}\!P+{}^{{}_v}\!P+{}^{{}_t}\!P
+\xi\, {}^{{}_t}\!\phi^{\mu\nu}\,{}^{{}_v}\!\phi_{\mu}\,j_\nu,
\label{svt model}
\end{equation}
where the shear and dilatation currents are exchanged between the vector and tensor fluids while the spin current is shared by all the types. Using Eqs. (\ref{spinor})-(\ref{tensor}) and field equations, we reach $\partial_\nu ({}^{{}_s}\!S^{\rho\sigma\cdot\nu}+{}^{{}_v}\!S^{\rho\sigma\cdot\nu}+{}^{{}_t}\!S^{\rho\sigma\cdot\nu})=0$.

\section{Local Symmetry}\label{LOCAL} 
So far, our discussions remain in the global symmetry under the transformation (\ref{global trans. x}) in a flat spacetime $R_4$. Now we are prepared to go for the local symmetry on the curved spacetime $(L_4,g)$. Local symmetry under $Di\!f\!f(4,R)$ requires the derivative to be covariantly modified; providing $\phi^I$ is an orthonormal-frame representation, it means
$\partial_\mu\phi^I\rightarrow \nabla_\mu\phi^I\equiv(\partial_\mu+\omega_\mu{}^a{}_b \mathsfbi{\Sigma}_a{}^b)\phi^I$, where the spin connection $\omega_{\mu\cdot ab}$ is introduced as a gauge potential. Now the orthonormal frame $e^\mu_a$ and its conjugate (coframe) $e^a_\mu$ ($e^\mu_ae^a_\nu=\delta^\mu_\nu$) vary in the spacetime. Unlike the Poincar$\acute{\textrm{e}}$-gauge theory, our generator $\mathsfbi{\Sigma}_a{}^b$ could contain the symmetric part, so the local-gauge symmetry requires $\omega_{\mu.ab}\neq-\omega_{\mu.ba}$, which leads to
\begin{subequations}
\begin{align}
\omega_\mu{}^a{}_b&=e^\rho_b e_\sigma^a\Gamma_{\mu\rho}{}^\sigma-e^a_{\rho,\mu}e^\rho_b,\label{spin connection}\\
\Gamma_{\alpha\beta\cdot \gamma}&
=\frac{1}{2}\Delta_{\alpha\beta\gamma}^{\mu\nu\rho}
(\partial_\mu g_{\nu\rho}-T_{\mu\nu\cdot\rho}+Q_{\mu\cdot\nu\rho}),\label{connection}
\end{align}
\end{subequations}
where $\Delta_{\alpha\beta\gamma}^{\mu\nu\rho}\equiv\delta^\mu_\alpha\delta^\nu_\beta\delta^\rho_\gamma+\delta^\mu_\beta\delta^\nu_\gamma\delta^\rho_\alpha-\delta^\mu_\gamma\delta^\nu_\beta\delta^\rho_\alpha$ \cite{Hehl76-rev}. Now $T_{\mu\nu\cdot\rho}(\equiv 2\Gamma_{[\mu\nu]\cdot\rho})$ and $Q_{\mu\cdot\nu\rho}(\equiv -\nabla_\mu g_{\nu\rho})$ are identified as the torsion and non-metricity. The fluid sector reads
\begin{equation}
S_\mathcal{F}=\int P(\mathfrak{D}\phi^I,\mathcal{W}) \sqrt{-g}d^4x,\label{hyperfluid 2}
\end{equation}
where $\mathfrak{D}\equiv w^\mu \nabla_\mu$ and $\mathcal{W}\equiv-\frac{1}{2}w^\mu w^\nu g_{\mu\nu}$. Following Refs. \cite{Obu14,Obu15}, we take $g_{\mu\nu}$ and $\Gamma_{\mu\nu\cdot \rho}$ as the independent gauge potentials whose variations yield the following coupling charges \cite{MAG}:
\begin{subequations}
\begin{align}
%\sqrt{-g}T^\mu_a&\equiv\frac{\delta S_\mathcal{F}}{\delta e^a_\mu}=\sqrt{-g}\left[(\rho+P)u^\mu u_a-Pe^\mu_a\right]\label{e eq.}\\
\!\!\!\!\sqrt{-g}\Delta^{\mu\nu\cdot \rho}&\equiv\!-\frac{\delta S_\mathcal{F}}{\delta\Gamma_{\rho\nu\cdot\mu}}\!=\!-\sqrt{-g}|\mathbf{w}|\sigma_I(\Sigma^{\nu\mu}){}^I{}_J\phi^J u^\rho,\label{Gamma eq.}\\
\sqrt{-g}\Theta^{\mu\nu}&\equiv-2\frac{\delta S_\mathcal{F}}{\delta g_{\mu\nu}}=\sqrt{-g}(T^{\mu\nu}+\overset{*}{\nabla}_\rho\Delta^{\mu\nu\cdot \rho}),\label{g eq.}
\end{align}
\end{subequations}
where $\overset{*}{\nabla}_\mu=\nabla_\mu-(\Gamma_{\rho\mu}{}^\rho-\tilde{\Gamma}_{\rho\mu}{}^\rho)$ ($\tilde{\Gamma}$: Levi-Civita connection). These charges are to be coupled with each field strength of the gravity sector. The variations by the fluid's variables $(\boldsymbol{\phi},\boldsymbol{w})$ read
\begin{subequations}
\begin{align}
&\frac{\delta S_\mathcal{F}}{\delta\phi^I}=-\sqrt{-g}\overset{*}{\nabla}_\rho\left(\sigma_I w^{\rho}\right)=0,\label{phi eq. 2}\\
&\frac{\delta S_\mathcal{F}}{\delta w^{\mu}}
=\sqrt{-g}(\sigma_I \nabla_\mu\phi^I-\tau w_{\mu})=0.
\label{w eq. 2}
\end{align}
\end{subequations}
which describe all the properties of our hyperfluid, while more intuitive picture may be obtained from the Euler equation; taking the gradient of $P(\Upsilon,\mathcal{W})$ and using Eqs. (\ref{phi eq. 2})-(\ref{w eq. 2}) yield
\begin{equation}
(\delta_\mu^\rho\overset{*}{\nabla}_\nu-T_{\mu\nu}{}^\rho)\{(\rho+P)u^\nu u_\rho\}=P_{,\mu}-R_{\rho\sigma\mu\nu}\,S^{\rho\sigma\cdot\nu},
\label{M-forced Euler eq.}
\end{equation}
where the Riemann-Christoffel curvature is defined in this paper by $R^\rho{}_{\sigma\mu\nu}\equiv 2\partial_{[\mu}\Gamma_{\nu]\sigma}{}^{\rho}+2\Gamma_{[\mu|\lambda}{}^{\rho}\Gamma_{|\nu]\sigma}{}^\lambda$. In contrast to the conventional models, Eq. (\ref{M-forced Euler eq.}) expresses very simple dynamics; only the Mathisson-Papapetrou force appears as the modification from the intrinsic hypermomentum, which is solely because the action (\ref{hyperfluid 2}) only contains the minimal coupling of $Di\!f\!f(4,R)$ gauge and free from any external constraints, providing an example of pure diffeomorphism-gauge theory. Further gauge coupling can be immediately incorporated. Local $SU(n)\times Di\!f\!f(4,R)$ symmetry imposed on $\boldsymbol{\phi}^I$ requires the covariant derivative $\mathcal{D}_\mu\equiv\partial_\mu+\omega_\mu{}^a{}_b \mathsfbi{\Sigma}_a{}^b-ie\mathsfbi{t}_aA^a_\mu$. The Lagrangian density $P(\Upsilon,\mathcal{W})$ ($\Upsilon \equiv(\mathfrak{D}\boldsymbol{\phi}_I)^\dagger\mathfrak{D}\boldsymbol{\phi}^I$, $\mathfrak{D}\equiv w^\mu\mathcal{D}_\mu$) results in 
\begin{equation}
(\delta_\mu^\rho\overset{*}{\nabla}_\nu-T_{\mu\nu}{}^\rho)\{(\rho+P)u^\nu u_\rho\}=P_{,\mu}-R_{\rho\sigma\mu\nu}\,S^{\rho\sigma\cdot\nu}-F^a_{\mu\nu}J_a^\nu,
\label{FM-forced Euler eq.}
\end{equation}
where $J^\nu_a$ and $S^{\rho\sigma\cdot\nu}$ are charge and spin currents given by Eqs.(\ref{charge current})-(\ref{charged-hypermomentum current}). Likewise, the present formalism always yields the Euler equation purely coupling with the gauge forces of corresponding symmetries, which is exactly due to the entire release from external constraints. 

Finally we shall mention another gauge-gravity formalism based on $GA(4,R)$ \cite{Hehl95}, where general transformation in the tangent bundle is taken as further gauging property, allowing the active interpretation of local translation, spin, shear, and dilatation of the frame. The vector and tensor fluids can be successfully formulated in $GA(4,R)$ framework leading to an identical set of equations to $Di\!f\!f(4,R)$-based theory, while the spinor fluid should be generalized by replacing the spinor with the spinor manifield \cite{Hehl95,Neeman77}. \\

%alternative formulation of the metric-affine gravity recently developed by \cite{Obu14,Obu15}, where the gravitational-field equations result from the variations of only the metric and affine connection.\\

\section{Conclusion} In this paper, hyperfluid was defined by its own right with its action free from any external constraints, regarding the hyperfluid as a self-consistent classical field. Discarding the phenomenological interpretations based on any underlying systems, the intrinsic hypermomentum is no more an additional variable accompanied by Lagrangian multipliers, but it naturally arises from the diffeomorphism covariance, which defines a pure diffeomorphism-gauge theory of the perfect fluid. It should be emphasized as a remarkable advantage that \emph{we only need to solve $\phi^I$ and $w^\mu$}, and the other fluid's properties such as the hypermomentum and charge currents are all calculated from them through Eqs. (\ref{hypermomentum current}), (\ref{charge current}), and (\ref{charged-hypermomentum current}). It is worth comparing the present formalism with those of Refs. \cite{KN14,Nair12,Jackiw04}, where Lie-group elements of required symmetries perform as their field variables; each conservation law requires elements of associated symmetry group. In contrast, $\phi^I$ in the present formalism belongs to the homogeneous space of the corresponding group on which we impose all the group symmetries at once. Indeed, we have shown that local $SU(n)\times Di\!f\!f(4,R)$ symmetry imposed on $\boldsymbol{\phi}^I$ successfully yields the gauge theory of the charged hyperfluid in Eq. (\ref{FM-forced Euler eq.}). Also we shall notice that the group symmetry alone does not specify the hyperfluid, but each representation $\phi^I$ of the same group symmetry results in a different hyperfluid, such as the scalar, spinor \cite{spinor}, vector, and tensor fluids, each of which has its possible class of interactions. In this classification, the ordinary fluid (the scalar fluid) can be understood as \emph{the hyperfluid with no hypermomentum}; our hyperfluid model gives a natural generalization of known fluid concept.

%These classifications of hyperfluids may be one of distinct features of the present formalism.\\ 
 
%we should notice that the \emph{chirality} can be introduced as a unique property of the spinor fluid; in the chiral representation we write ${}^{{}_s}\!\phi=[{}^{{}_s}\!\phi_L,{}^{{}_s}\!\phi_R]{}^T$, with which one can introduce the interaction breaking the chiral symmetry via an axial vector $\phi$. These classifications in hyperfluid do not appear in the frameworks of \cite{Nair12,KN14}.
% The present model gives the simplest construction of the hyperfluid which only contains $\phi^I$ and $w^\mu$ as its dynamical variables.
%It is also remarkable that the S-fluid defines, by itself, the spin fluid without imposing the constraint on the dilatation and shear currents. 

Release from the external constraints enables the intrinsic hypermomentum to play more active role in interactions; $SU(n)$-gauge theory, interaction between fluids, and the metric-affine gauge theory were examined where hypermomentum exchange is successfully described on the basis of the field-theoretic interactions, suggesting that the hyperfluid can be studied as a self-consistent field at least in the classical framework. Free from the phenomenology trivially imposed by any external constraints, non-trivial phenomena may result from our self-consistent formalism. Indeed, regarding the charged hyperfluid, we have shown a possible model of non-minimal coupling using the $SU(n)$-magnetic moment $\chi^a_{\mu\nu}$. Note that the magnetic moment $\chi^a_{\mu\nu}$ of Eq. (\ref{magnetic moment}) is not constrained to the spin density $\Delta_{[\mu\nu].0}$ and is even able to exchange the dilatation and shear currents (see Eq. (\ref{hypermomentum non-conversion})). This may be understood as further generalization of spin-orbit interaction, where intrinsic dilatation and shear currents may cause dilatational and straining flow. 

The removal of constraints further enhances the potential of the theory. It has been recently shown by Ref. \cite{AM16} that the natural symplectic structure is obtained by discarding the Lin constraint. Such a fundamental modification at the classical stage offers the first step to the canonical QFT of fluid in the generalized spacetime geometry.\\

\subsection*{Acknowledgement}
The author currently belongs to Institute of Materials and Systems for Sustainability, Nagoya University, Japan. The present work is supported by the Japan Society for the Promotion of Science for Young Scientists.\\


\begin{thebibliography}{99}
\bibitem{frame}
Note that the orthonormal frame $e^\mu_a$ here is defined as global coefficients, as is clear from the fact that the metric $g_{\mu\nu}$ now globally transforms according to Eq. (\ref{global trans. x}).
\bibitem{MAG}
In contrast to the $GA(4,R)$-gauge theory, $Di\!f\!f(4,R)$-based theory does not take the variation in coframe $e^a_\mu$ whose Euler-Lagrange equation is missing. The lost equation, however, can be reproduced from the remaining equations as far as we employ the standard gravity sector in the $GA(4,R)$-gauge theory, which was recently proven by \cite{Obu14}.
\bibitem{spinor}
Strictly speaking, the spinor fluid, based on the 4-spinor field, relates to the Lorentz group through its double covering, while the theory acquires the diffeomorphism covariance by the frame field.
\end{thebibliography}
\end{document}